\documentstyle[epsfig,psfig,referee]{mn}

\def\tN{{\widetilde N}}
\def\tA{{\widetilde A}}

\def\td{{\tilde d}}
\def\ts{{\tilde s}}

\def\teps{{\tilde\varepsilon}}
\def\cF{{\cal F}}
\def\cG{{\cal G}}
\def\bN{{\bar N}}
\def\bT{{\bar T}}
\def\pdot{{\dot \phi}}
\def\ie{{\it i.e.}}
\def\eg{{\it e.g.}}
\begin{document}
\title[CMB Noise \& Signal Estimation]{Simultaneous Estimation of Noise and Signal  in Cosmic Microwave Background
  Experiments }
\author[P.~G.~Ferreira, A.~H.~Jaffe]
{P.~G.~Ferreira$^{1,2}$\thanks{\tt pgf@mail.cern.ch}, A.~H.~Jaffe$^{3}$
\thanks{\tt jaffe@cfpa.berkeley.edu}\\
$^{1}$ Theory Group, CERN, CH-1211, Geneve 23, Switzerland\\
$^{2}$ CENTRA, Instituto Superior Tecnico, Lisboa 1096 Codex, Portugal\\
$^{3}$ Center for Particle Astrophysics,
  301 LeConte Hall, University of California, Berkeley, CA 94720, USA}
\maketitle

\begin{abstract}
  To correctly analyse data sets from current microwave detection
  technology, one is forced to estimate the sky signal and experimental
  noise simultaneously. Given a time-ordered data set we propose a
  formalism and method for estimating the signal and associated errors
  without prior knowledge of the noise power spectrum. We derive the
  method using a Bayesian formalism and relate it to the standard
  methods; in particular we show how this leads to a change in the
  estimate of the noise covariance matrix of the sky signal. We study
  the convergence and accuracy of the method on two mock observational
  strategies and discuss its application to a currently-favoured
  calibration procedure.
 \end{abstract}


\section{Intro}

Observations of temperature fluctuations in the Cosmic Microwave
Background (CMB) are poised to become perhaps the most precise tools to
probe cosmological models (\eg, Bond \& Jaffe 1998). 
Characteristics of microwave detection
technology---in particular, long-time-scale noise correlations---make
these observations especially challenging 
\cite{delabrouille,debernardis,wgh,tegmapsa}.  Such difficult
observations---and such high scientific stakes---require equally careful
analyses of the data. A map of the CMB sky is not enough; detector
characteristics will ensure that the noise contribution to such a map is
highly correlated from pixel to pixel, in a pattern depending in detail
upon the detector noise spectrum and the observing strategy. But correct
estimation of the map and the noise is especially crucial for
cosmological analyses which are themselves trying to measure a variance,
the intrinsic power spectrum, $C_\ell$ of the CMB---mis-estimation of
the noise can directly add or subtract (\ie, bias) the estimated
$C_\ell$ \cite{gorski94a,gorski94b,lineweaver}.

Unfortunately, we usually do not have an {\em independent} determination
of the detector noise power spectrum, at least not one that has been
made under the same operating conditions as the actual experiment.
Moreover, there are noise contributions such as atmospheric emission
which may have similar characteristics but are completely independent of
the instrument. For this reason, we prefer to estimate the noise
directly from the experimental data, perhaps using prior estimates of
the noise under different conditions as a guide. In this paper, we
discuss how to jointly estimate the microwave sky map and the noise
correlation matrix using a maximum-likelihood formalism. This brings
together two strands of CMB data analysis that had previously been seen
as separate steps. We also discuss the further use of the map and
correlation matrix for power spectrum estimation, and the pitfalls that
this may incur relative to the ideal (but impossible to implement!) full
analysis. Although CMB experiments have had to address this problem in
the past, only cursory descriptions have been made of the
methods used to estimate the noise and the validity of the
approximations \cite{lineweaver,qmap}.

In Section~\ref{sec:formalism} we begin with a discussion of CMB
observations and a review of the Bayesian formalism for CMB mapmaking
and parameter estimation. We extend this formalism to new maximum
likelihood estimates for both the map and noise correlations. In
Section~\ref{sec:method} we discuss an iterative method for determining
this maximum likelihood and apply it to a two model observational
strategies as a test. In
Section~\ref{sec:dipole} we present the problem of estimating the noise
for a data stream with large signal-to-noise, namely calibration 
with the dipole.
In Section~\ref{sec:conclusions} we conclude.

\section{Bayesian Formalism}
\label{sec:formalism}

As has become customary, we start our analysis with Bayes' theorem
\begin{equation}
  \label{eq:bayestheorem}
  P(\theta|DI) \propto P(\theta|I) P(D|\theta I)
\end{equation}
where $\theta$ are the parameters we are trying to determine, $D$ is the
data, and $I$ is the ``background information'' describing the problem
\cite{lupton}, which we will often omit from our probability
expressions. The quantity $P(D|\theta I)$ is thus the likelihood, the
probability of the data given a specific set of parameters,
$P(\theta|I)$ is the prior probability for the parameters, and the
left-hand side of the equation is the posterior probability for the
parameters given the data.

Here, we will take the data, $d_i$, as given by a time series of CMB
measurements, 
\begin{equation}
  \label{eq:timeseries}
  d_i = s_i + n_i = \sum_p A_{ip} T_p + n_i,
\end{equation}
where $i$ labels the time, $t=i \ \delta t$, $s_i$ and $n_i$ are the
experimental noise and sky signal contributions at that time. The signal
is in turn given by the operation of a ``pointing matrix,'' $A_{ip}$ on
the sky signal at pixel $p$, $T_p$ (\ie, the ``map''); we take the
latter to be already pixelized and smeared by the experimental beam, so
$A$ is a very sparse matrix with a single ``1'' entry for each time
corresponding to the observed pixel. In what follows we will often rely on 
the summation convention and write $\sum_p A_{ip} T_p = A_{ip}T_p$, or
occasionally use matrix notation, as in $AT$, etc.

We will assume that the observed noise $n_i$ is a realization of a
stationary Gaussian process with power spectrum $\tN(\omega)$. This means
that the correlation matrix of the noise is given by
\begin{equation}
  \label{eq:noisecorrelation}
  \langle n_i n_{i'} \rangle = N_{ii'} = \int {d\omega\over2\pi}\;
  \tN(\omega) e^{-i\omega (t_i-t_{i'})}.
\end{equation}
The stationarity of the process requires (or is defined by) $N_{ii'} =
N(t_i-t_{i'})$.

Most generally, we will take the parameters to be
\begin{itemize}
\item The observed CMB signal on the sky, $T_p$;
\item The power spectrum of the noise, $\tN(\omega)$
\item (Possibly) any cosmological parameters which describe the
  distribution of the $T_p$ (\ie, the CMB power spectrum, $C_\ell$,
  although we could also directly use the cosmological parameters such as $H_0$
  and $\Omega$).
\end{itemize}
Sometimes, we will assign a prior distribution for the CMB signal on the
sky itself ({\em i.e.}, uniform in the sky temperature) such that the
cosmological parameters (or $C_\ell$) will be irrelevant---we will see
that this doing this as an intermediate step retains all of the
information in the dataset. At other times we will marginalize over the
CMB signal itself and determine those parameters.

With these parameters and the data, $d_i$, Bayes' theorem becomes
\begin{eqnarray}
  \label{eq:bayescmb}
  P[T_p,\tN(\omega),C_\ell | d_i,I] &\propto& 
  P[\tN(\omega)|I] P(T_p,C_\ell | I) \nonumber \\
  && \times P[d_i | \tN(\omega), T_p, I].
\end{eqnarray}
Here, we have used two pieces of information to simplify slightly.
First, the noise power spectrum, $\tN(\omega)$ does not depend at all on
the signal, so we can separate out its prior distribution.\footnote{In
  principle, the noise may depend on the signal such as in the case of a
  nonlinear response of the system to large-amplitude signals such as
  planets or the galaxy; in practice data thus contaminated is
  discarded.}
Second, given the noise power spectrum and the sky
signal, the likelihood does not depend upon the cosmological parameters.
For the Gaussian noise we assume, the likelihood is simply
\begin{eqnarray}
  \label{eq:noiselike}
  -2\ln{\cal L}&\equiv&-2\ln P[d_i | \tN(\omega), T_p]  \nonumber\\
  &=& \ln \left| N_{ii'} \right| +
  \sum_{ii'} (d_i-s_i)N^{-1}_{ii'}(d_{i'}-s_{i'}) \nonumber\\
  &=& \sum_k \left[ \ln \tN_k +
    |\td_k-\ts_k|^2/\tN_k \right]
\end{eqnarray}
(ignoring an additive constant); recall that $s_i=\sum_p A_{ip}T_p$.
The second equality uses tildes to denote the discrete Fourier transform
at angular frequency $\omega_k$.

We will now apply these general formulae to various cases.

\subsection{Known noise power spectrum}
\label{sec:knownnoise}
We will start with the simplest case, where we have complete prior
knowledge of the noise power spectrum. This is the case that has been
previously discussed in the literature, but we emphasize that it is very
unrealistic.

We assign a delta-function prior distribution to $N$, transforming it in
effect from a parameter to part of the prior knowledge. First, we assume
no cosmological information about the distribution of temperatures on
the sky: $P(T_p, C_\ell | I ) = P(T_p | I) P(C_\ell | I)$; with this
separation the posterior for $C_\ell$ is simply the prior---the
experiment gives us no new information. We will also assign a uniform
prior to $T_p$, lacking further information. Now, the posterior
distribution for the sky temperature is simply proportional to the
likelihood, which can be rewritten by completing the square in the
exponential as
\begin{eqnarray}
  \label{eq:maplike}
  P(T_p | \tN, d, I) &\propto& P(d|T_p,\tN,I) \nonumber\\
&\propto&
 {1\over\left| 2\pi C_{Npp'} \right|^{1/2}}\nonumber\\&&\times
\exp\left[
-{1\over2} \sum_{pp'} \left(T_p - \bT_p\right) C^{-1}_{N,pp'}
    \left(T_{p'} - \bT_{p'}\right)\right]
\end{eqnarray}
with the mean (also, the likelihood maximum) given by
\begin{equation}
  \label{eq:lsmap}
  \bT = \left( A^T N^{-1} A\right)^{-1} A^T N^{-1} d
\end{equation}
(in matrix notation), and the noise correlation matrix by
\begin{equation}
  \label{eq:noisecorr}
  C_N = \left( A^T N^{-1} A\right)^{-1}.
\end{equation}
Occasionally, the inverse of this correlation matrix is referred to as
the {\em weight matrix}.
As is usual for linear Gaussian models, the mean is just the
multidimensional least-squares solution to $d = A T$ with noise
correlation $N$.  This is just the standard mapmaking procedure 
\cite{lupton,tegmapsb,wrightprocedure}, cast into the form of a
Bayesian parameter-estimation problem.

For the case of known noise, however, this map is more than a just a
visual representation of the data. Even if we wish to determine the
cosmological parameters, it is an essential quantity: we can write
the prior for both the map and the spectrum as
\begin{equation}
  \label{eq:sigprior}
  P(C_\ell, T_p | I) = P(T_p | C_\ell, I) P(C_\ell | I)
\end{equation}
using the laws of probability, and so we can see that our rewriting of
the likelihood in the form of Eq.~\ref{eq:maplike} remains useful. That
is, the full distribution is only a function of the data through the
maximum-likelihood map, $\bT$---in statistical parlance, $\bT$
is a {\em sufficient statistic}.  Thus for known noise, we can {\em
  always} start by making a map (and calculating its noise matrix,
$C_N$).

\subsection{Cosmological CMB priors}
\label{sec:cosmopriors}
Here, we briefly examine the specific form of the signal prior, $P(T_p |
C_\ell, I)$, motivated by simple Gaussian models. That is, we take the
sky temperature, $T_p$, to be an actual realization of a Gaussian CMB
sky, with covariance specified by the power spectrum, $C_\ell$,
\begin{equation}
  \label{eq:signalcovar}
  \langle T_p T_{p'} \rangle = C_{T,pp'} =
  \sum_\ell\frac{2\ell+1}{4\pi}C_\ell B_\ell^2 P_\ell
  \left({\hat x}_p\cdot{\hat x}_{p'}\right)
\end{equation}
(note that we include beam-smearing by a symmetric beam with spherical
harmonic transform $B_\ell$ in this definition); ${\hat x}_p\cdot{\hat
  x}_{p'}$ gives the cosine of the angle between the pixels. With this
covariance, the prior becomes
\begin{equation}
  \label{eq:signalprior}
  P(T_p |C_\ell, I) = {1\over\left| 2\pi C_{Tpp'} \right|^{1/2}}
  \exp\left[ -{1\over2} \sum_{pp'} T_p C^{-1}_{T,pp'} T_{p'}\right].
\end{equation}
We thus have a posterior distribution for $T_p$ and $C_\ell$ which is
the product of two Gaussians, $P(C_\ell,T_p|d,I)\propto P(T_p|C_\ell,I)
P(d|T_p,I)$, given by Eqns.~\ref{eq:maplike} and \ref{eq:signalprior}.
In the usual cosmological likelihood
problem, we don't care about the actual sky temperature {\em per se},
but are concerned with the $C_\ell$ (or the parameters upon which the
power spectrum depends). Thus, we can marginalize over the $T_p$,
\begin{eqnarray}
  \label{eq:mapmarginalize}
  P(C_\ell | d, I) &=& \int dT_p\; P(C_\ell,T_p|d,I) = 
P(C_\ell|I) \int dT_p\; P(T_p|C_\ell,I) P(d|T_p,I) \nonumber\\ &=&
P(C_\ell|I) P(\bT(d)|C_\ell I),\nonumber\\
&=& P(C_\ell|I) \int dT_p\;      
{1\over \left| 2\pi C_T \right|^{1/2}}  {1\over \left| 2\pi C_N \right|^{1/2}}
\nonumber\\
&&\qquad\times\exp\bigg[-{1\over2} \sum_{pp'}
\left( T_p C^{-1}_{T,pp'} T_{p'} +
      \left(T_p - \bT_p\right) C^{-1}_{N,pp'} 
      \left(T_{p'} - \bT_{p'}\right) \right)\bigg], \label{theq}
\end{eqnarray}
where we have included a prior for the power spectrum itself, so we can
write the Gaussian factor as the likelihood for the map given $C_\ell$,
$ P(C_\ell | d, I)\propto P(C_\ell|I)P(\bT|C_\ell)$.
Equation \ref{theq} defines the effective likelihood for the map ($\bT$, now
considered as the data rather than the timestream itself, $d$), which is
easily computed again by completing the square, giving
\begin{equation}
  \label{eq:maplikelihood}
  P(\bT_p | C_\ell I) =
  {1\over\left| 2\pi\left(C_{Tpp'} + C_{Npp'}\right)\right|^{1/2}}
\exp\left[ -{1\over2} \sum_{pp'} \bT_p \left(C_T+C_N\right)^{-1}_{pp'}
    \bT_{p'}\right].
\end{equation}
This is just the usual CMB likelihood formula: the ``observed map,''
$\bT_p$, is just the sum of two quantities (noise and signal)
distributed as independent Gaussians. Note again that the data 
only enter through the maximum likelihood map, $\bT$, although that
calculation is only implicit in this formula. Further, the power
spectrum $C_\ell$ only enters through the signal correlation matrix,
$C_T$, and in a very nonlinear way.

We can also play a slightly different game with the likelihood. If we
retain the Gaussian prior for the CMB temperature but {\em fix} the CMB
power spectrum, we can estimate the map with this additional prior
knowledge. We will again be able to complete the square in the
exponential and see that $T_p$ is distributed as a Gaussian:
\begin{equation}
  \label{eq:wienerlike}
  P(T_p | C_\ell, d, I)={1\over \left| 2\pi C_W \right|^{1/2}} 
  \exp\left[-{1\over2} \chi^2(T_p | C_\ell, d, I)\right]
\end{equation}
with
\begin{equation}
  \chi^2(T_p | C_\ell, d, I)\equiv\sum_{pp'} \
      \left(\bT_p - (W\bT)_p\right) C^{-1}_{W,pp'} 
      \left(\bT_{p'} - (W\bT)_{p'}\right).
\end{equation}
Now, the mean is given by
\begin{equation}
  \label{eq:wiener}
  (W\bT) = C_T(C_T + C_N)^{-1}\bT
\end{equation}
which is just the {\em Wiener Filter}. It has correlation matrix given
by
\begin{equation}
  \label{eq:wienercorr}
  C_W = C_T(C_T + C_N)^{-1}C_T.
\end{equation}
Note that the maximum-likelihood map, $\bT$ still appears in these
formulae, but it is no longer the maximum of the {\em posterior}
distribution, now given by the Wiener filter, $W\bar T$.

This subsection has shown how many of the usual CMB data calculations
can be seen as different uses of the Bayesian formalism:
\begin{itemize}
\item the least-squares map (seeing it as a ``sufficient statistic''),
\item the CMB cosmological-parameter likelihood function, and
\item the Wiener filter map of the CMB signal.
\end{itemize}
The differences depend on what quantity is estimated (the map or the
power spectrum) and what prior information is included.

\subsection{Unknown noise}
\label{unknownnoise}
The previous subsection briefly outlined the Bayesian approach to CMB
statistics, assuming a known noise power spectrum. Now, we will relax
this assumption and approach the more realistic case when we must
estimate both the experimental noise and the anisotropy of the CMB. We
will take as our model a noise power spectrum of amplitude $\bN_\alpha$
in bands numbered $\alpha$, with a shape in each band given by a fixed
function $P_k$ with a width of $n_\alpha$; 
$n_\alpha=\frac{\Delta_\omega}{\omega_0}$ is the number
of discrete modes in the band $\alpha$ where $\Delta_\omega$ is the width
of the band and $\omega_0$ is the minimum frequency of the data-stream. 
We will usually take $P_k={\rm const}$ so $\tN$ is
piecewise constant. That is,
\begin{equation}
  \label{eq:noisemodel}
  \tN(\omega_k) = \bN_\alpha P_k.
\end{equation}

We again assign a constant prior to the sky map, $T_p$. As the prior for
the noise we will take $P(\bN_\alpha)\propto1/\bN_\alpha^\nu$. 
With $\nu=1$ and a single band, this is the usual Jeffereys prior
advocated for ``scale parameters''  and the units on ${\bar
  N}_\alpha$ are irrelevant \cite{kso}. 

With this model and priors, the joint likelihood for the noise and the
map becomes
\begin{eqnarray}
  \label{eq:jointprob}
  P(T_p, \bN_\alpha | d,I) &\propto& \prod_\alpha
  {1\over\bN_\alpha^{\nu+n_\alpha/2}} 
  \exp\left[ -{1\over2\bN_\alpha}
    \sum_{k\in\alpha}\frac{1}{P_k}\left|\td_k - \tA_{kp}T_p\right|^2\right]
\nonumber\\
  &=&  \prod_\alpha
  {1\over\bN_\alpha^{\nu+n_\alpha/2}} 
  \exp\left[ -{1\over2\bN_\alpha}
    \sum_{k\in\alpha}\frac{1}{P_k}\left|\teps_k\right|^2\right]
\end{eqnarray} 
where ${k\in\alpha}$ refers to a sum over modes in band $\alpha$. We
also define the estimate of the noise as $\varepsilon=d-AT$ for future use.

We can simultaneously solve for the maximum-probability noise and
signal. Carrying out the necessary derivatives gives
\begin{eqnarray}
  \label{eq:Tpderiv}
  {\partial \ln{\cal L}\over\partial T_p} &=& \sum_\alpha\left[
    {1\over\bN_\alpha}\sum_{k\in\alpha} \frac{1}{P_k}
    \left(\td_k-\tA_{kp'}T_{p'}\right)\tA_{kp}\right]\nonumber\\
&=& (d-AT)^T N^{-1} A = \varepsilon^T N^{-1} A
\end{eqnarray}
(switching between indices in Fourier space and matrix notation) and 
\begin{equation}
  \label{eq:Nderiv}
  {\partial \ln{\cal L}\over\partial \bN_\alpha} = -{1\over2}
  {1\over\bN_\alpha}\left[(n_\alpha+2\nu)-{1\over\bN_\alpha}
      \sum_{k\in\alpha} 
      \frac{1}{P_k}\left|\td_k-\tA_{kp'}T_{p'}\right|^2\right].
\end{equation}
Setting these to zero and solving, we then find we must simultaneously
satisfy
\begin{equation}
  \label{eq:maxLmap}
  T = (A^T N^{-1} A)^{-1} A^T N^{-1} d
\end{equation}
(using matrix notation for simplicity) and
\begin{equation}
  \label{eq:maxLnoise}
  \bN_\alpha = {1\over n_\alpha+2\nu}\sum_{k\in\alpha}
  \frac{1}{P_k}\left|\td_k-\tA_{kp'}T_{p'}\right|^2 = 
 {1\over n_\alpha+2\nu}\sum_{k\in\alpha}
  \frac{1}{P_k}\left|\teps_k\right|^2.
\end{equation}
Equation~\ref{eq:maxLmap} is just the usual maximum-likelihood map
solution; Equation~\ref{eq:maxLnoise} is the average ``periodogram''
(\ie, the naive power spectrum calculated from the Fourier transform) of the 
noise over the band $\alpha$, with a slight modification for the prior
probability, parameterized by $\nu$; for wide bands this modification 
is irrelevant.
As we will see below, iteration is actually a very efficient way to solve 
Equations \ref{eq:maxLmap} and \ref{eq:maxLnoise} 
simultaneously. For future reference, we also write down the derivative 
with respect to the map at the joint maximum (\ie, substituting
Eq.~\ref{eq:maxLnoise} into Eq.~\ref{eq:Tpderiv}),
\begin{equation}
  \label{eq:maxLTpderiv}
    {\partial \ln{\cal L}\over\partial T_p} =\sum_\alpha\left[
      \left(n_\alpha+2\nu\right){
   \sum_{k\in\alpha} \teps_k\tA_{kp}/P_k
\over
   \sum_{k'\in\alpha} \left|\teps_{k'}\right|^2/P_{k'}}
      \right].
\end{equation}

If we fix the noise at the joint maximum, then the problem reduces to
that of the previous section, a Gaussian likelihood in $T_p$ for which
the usual tools can be applied. This is not a rigorous approach to the
problem, however due to correlation between noise and signal
estimation. To get a handle on this, we calculate the curvature of the
distribution around this joint maximum, which we define as
\begin{equation}
  \cF = \left(
    \begin{array}{cc}
      \cG_{pp'} & \cG_{p\alpha} \\
      \cG_{p\alpha} & \cG_{\alpha\alpha'} \\
    \end{array}\right)
\end{equation}
where a subscript $\alpha$ or $p$ refers to a derivative with respect to
$\bN_\alpha$ or $T_p$, respectively.  $\cF$ refers to the full matrix;
$\cal G$ to the sub-blocks. Explicitly, the parameter derivatives are
\begin{equation}
  \label{eq:maxcurvTT}
  {\partial^2 \ln{\cal L}\over\partial T_p\partial T_{p'}}=\cG_{pp'} = 
-\sum_\alpha \frac{1}{\bN_\alpha}\sum_{k\in\alpha}\frac{\tA_{kp}\tA_{kp'}}{P_k}
=-A^T N^{-1} A
\end{equation}
and
\begin{equation}
  \label{eq:maxcurvNN}
  {\partial^2 \ln{\cal L}\over\partial \bN_\alpha\partial
    \bN_{\alpha'}}=
  \cG_{\alpha\alpha'}=-\frac{n_\alpha+2\nu}{2\bN_\alpha^2}
  \delta_{\alpha\alpha'}
\end{equation}
with the cross-curvature
\begin{equation}
  \label{eq:maxcurvTN}
  {\partial^2 \ln{\cal L}\over\partial T_p\partial \bN_{\alpha}}=
\cG_{p\alpha}=-\frac{1}{\bN_\alpha^2}
\sum_{k\in\alpha}\frac{\teps_k \tA_{kp}}{P_k}.
\end{equation}

The distribution about the joint maximum is {\em not} a Gaussian in the
$\bN_\alpha$ directions, so there is more information available than
this. Nonetheless, if we treat the distribution as if it were Gaussian,
we can calculate the covariance matrix, given by the inverse of this
curvature matrix; if we assume that the matrix 
$I_{pp'}-\cG_{pp''}^{-1}\cG_{p''\alpha}\cG_{\alpha\alpha'}^{-1}\cG_{\alpha'p'}$
is invertible, then we find that
 the effective variance of the map is increased from $C_{Npp'}=\cG_{pp'}^{-1}$
to 
\begin{eqnarray}\label{eq:effectivevar}
C^{\rm eff}_{N pp'} &=& (C_{N \ pp'}^{-1}-\cG_{p\alpha}\cG^{-1}_{\alpha\alpha'}\cG_{\alpha'p'})^{-1}\nonumber\\
 &=&\left [\sum_\alpha \frac{1}{\bN_\alpha}\sum_{k\in\alpha}\frac{\tA_{kp}\tA_{kp'}}{P_k}- 2\sum_\alpha \frac{1}{(n_\alpha + 2\nu)\bN^2_\alpha}
  \sum_{k\in\alpha}\frac{\teps_k\tA_{kp}}{P_k}
  \sum_{k'\in\alpha}\frac{\teps_{k'}\tA_{k'p'}}{P_{k'}}\right ]^{-1}
\label{correction}
\end{eqnarray}
which we have evaluated at the simultaneous peak of the distribution to
determine $\bN_\alpha$.

To assess the importance of the correction term we can 
consider an approximate expression that relates 
$(C_{N}^{\rm eff})^{-1}$ to $(C_{N})^{-1}$; we shall assume
that ${\tilde N}(\omega)$ is approximately white, with a constant
number of frequency samples per bin, \ie, $n_\alpha=n$, 
$P_k=1$ for all $\alpha$, and $\nu=1$. The key parameter is then
\begin{eqnarray}
r_p=\frac{4\pi}{\Delta_\omega\tau_p} 
\end{eqnarray}
where $\tau_p$ is the total time spent on a pixel $p$ and $\Delta_\omega$ is
the dimensionful width of the bins. A simple way of obtaining
 $r_p$ is by reordering the data stream in
such a way that all time samples for a given pixel are grouped contiguously.
Then the minumum frequency for a given pixel will correspond to 
$2\pi/\tau_p$ where $\tau_p$ is now the length of the  data stream segment.
We then have that the total number of discrete modes in bin of width 
$\Delta_\omega$ is $2\pi/{\Delta_\omega\tau_p}$. The extra factor of
two comes from equation (\ref{correction}).

 If we take the ensemble average
of Eq.~\ref{correction} we obtain
\begin{eqnarray}
\langle \left(C_{Npp'}^{\rm eff}\right)^{-1} \rangle=
\left(1-\sqrt{r_pr_{p'}}\right)
\sum_\alpha \langle\frac{1}{\bN_\alpha}\rangle
\sum_{k\in\alpha}\tA_{kp}\tA_{kp'}
=\left(1-\sqrt{r_pr_{p'}}\right)
\langle\left(C_{Npp'}\right)^{-1}\rangle.
\label{corrapp}
\end{eqnarray}
For an equal amount of time spent on each pixel we have that
$n=\Delta_\omega T_{\rm tot}/2\pi$ where $T_{tot}\equiv N_{\rm pix}\tau_p$
so that Eq.~\ref{corrapp} simplifies to
\begin{eqnarray}
\langle \left(C_{Npp'}^{\rm eff}\right)^{-1} \rangle
=\left(1-\frac{2N_{\rm pix}}{n}\right)
\langle\left(C_{Npp'}\right)^{-1}\rangle.
\end{eqnarray}
This immediately shows us two limits. In the case of one frequency per band,
the inverse variance is negative and the problem is ill-defined; in some
sense one is trying to estimate too much from the data set. In the limit of
just one band (\ie, $n\rightarrow\infty$) the correction for
the effective variance is negligible. For practical cases one needs
enough bins to properly capture the shape of the noise periodogram,
so one has to consider some intermediate regime. For the
case of one pixel and white noise, to get a correction of less
than 1$\%$ to the $C_{{\rm N}pp'}$, one needs $n>10^2N_{pix}$. 
For more realistic cases of many pixels and non-white noise, much 
larger bins are necessary, and must be evaluated on a case by case basis.

We note a caveat to this procedure: consider an experiment that simply
chops with a given frequency of rotation in a circle on the sky (as in a
simple model for Planck's observing strategy). Any source of noise
synchronous with this motion will necessarily be very difficult to
distinguish from sky signal. That is, there will necessarily be strong
covariance between chop- or spin- synchronous noise and the signal.
This will remain somewhat the case even if the pattern moves slowly on
the sky. In this case, other information, such as the smoothness of the
noise power spectrum, must be used.

\subsection{Noise Marginalization}

Instead of this joint solution, formally, at least, we know the
appropriate procedure: marginalize over the quantity we don't care about
(the noise power spectrum) to obtain the distribution for the quantity
we wish to know (the map, $T_p$). We can actually carry out the
integral in this case:
\begin{eqnarray}
  \label{eq:noisemarg}
  P(T_p | d_i,I) &=&  \int d\tN_k\; P(T_p, N_k| d_i, I) \nonumber\\
  &\propto& \prod_\alpha\left[\sum_{k\in\alpha}
    \frac{1}{P_k}\left|\td_k - \tA_{kp}T_p\right|^2\right]^{-(n_\alpha/2+\nu-1)}
\end{eqnarray}
(this is just Student's t distribution in a slightly different form than 
is usually seen).
If we stay in Fourier space, the maximum of this distribution can be
calculated to be the solution of
\begin{equation}
  \label{eq:margmax}
    {\partial \ln P \over\partial T_p} =\sum_\alpha\left[
       \left(n_\alpha+2\nu-2\right){
     \sum_{k\in\alpha} \frac{1}{P_k}\left(\td_k-\tA_{kp'}T_{p'}\right)\tA_{kp}
\over
     \sum_{k'\in\alpha}\frac{1} {P_{k'}}\left|\td_{k'}-\tA_{k'p'}T_{p'}\right|^2}
      \right].
\end{equation}
Note that this is {\em exactly the same form} as
Eq.~\ref{eq:maxLTpderiv}, the equation for the maximum probability
map in the joint estimation case, with the prefactor $(n_\alpha+2\nu)$
replaced by $(n_\alpha+2\nu-2)$. This
is equivalent to changing the exponent of the prior probability from
$\nu$ to $\nu-1$: the marginalized maximum for $\nu=1$ (Jefferys prior)
is the same as the joint maximum for $\nu=0$ (constant prior). We have
also seen that for $n_\alpha\gg1$, the value of $\nu$ is irrelevant, so
that these maxima should be nearly equal. (Moreover, the numerical tools to
calculate the joint solution, outlined below, can be used in this case,
too.) Note also that if $n_\alpha+2\nu-2\le 0$, the equation isn't solved
for any map. In this case, either our prior information is so
unrestrictive or the bands are so narrow that it is impossible to
distinguish between noise and signal, and the probability distribution
has no maximum when the noise is marginalized.

In principle, any further analysis of the map would have to rely on the
full distribution of Eq.~\ref{eq:noisemarg}. In practice, the
t-distribution is quite close to a Gaussian although the tails are 
suppressed by a power-law rather than an exponential. 
 It will thus be an excellent
approximation to take the distribution to be a Gaussian
\begin{eqnarray}
  \label{eq:noisemargapprox}
  P(T_p | d_i,I) &\approx& 
  {1\over\left| 2\pi C_{N,pp'}^{\rm eff} \right|^{1/2}}\nonumber\\
  &&\times
  \exp\left[ -{1\over2} \sum_{pp'} 
\left(T_p - {\bar T}_p\right) \left(C^{\rm eff}_{N,pp'}\right)^{-1}
    \left(T_{p'} - {\bar T}_{p'}\right)\right],
\end{eqnarray}
with noise covariance given by
\begin{eqnarray}
  \label{eq:CNeff}
  \left(C_{Npp'}^{\rm eff}\right)^{-1} &=& 
  -{\partial^2 \ln P(T_p | d_i,I)\over\partial
    T_p\partial T_{p'}}\nonumber\\
  &=&\sum_\alpha \left[\frac{(n_\alpha+2\nu-2)}{\sum_{k\in\alpha}|\teps_k|^2}\sum_{k\in\alpha}\frac{\tA_{kp}\tA_{kp'}}{P_k}- 2 \frac{(n_\alpha+2\nu-2)}
{(\sum_{k\in\alpha}|\teps_k|^2)^2 }
  \sum_{k\in\alpha}\frac{\teps_k\tA_{kp}}{P_k}
  \sum_{k'\in\alpha}\frac{\teps_{k'}\tA_{k'p'}}{P_{k'}}\right ],
\end{eqnarray}
where ${\bar T}$ is the solution to Eq.~\ref{eq:margmax} and the
derivatives are taken with respect to the {\em full} distribution of
Eq.~\ref{eq:margmax}.  Note that the inverse of $C_N^{\rm eff}$
(occasionally referred to as the weight matrix) is given by the sum of
two terms. The first is just the weight matrix that would be assigned
given the maximum probability solution for the noise,
Eq.~\ref{eq:noisecorr} (with the numerically irrelevant change of the
prior $\nu\to\nu-1$ as noted above). Again we find a correction term due
to the fact that we are only able to {\em estimate} this noise power
spectrum. This correction term can be minimized by taking sufficiently wide
bands, as in the simultaneous estimation case above (and subject to the
same caveats).

\section{An iterative method: convergence and accuracy}
\label{sec:method}
In this section we propose an algorithm for finding the Maximum Likelihood 
Estimate (MLE)
 of the noise and signal and we apply
it to two models that contain all the essential features of current
CMB measurement processes.
We parameterize the map and noise as described above, in terms of
a set of pixel temperatures, $T_p$, and a set of bandpowers, $\bN_\alpha$.
The noise is stationary and Gaussian with a periodogram of the form:
\begin{eqnarray}
\tN(\omega)\propto 1+\frac{\omega_{\rm knee}}{\omega} \label{eq:noiseper}.
\end{eqnarray}
Given some prior knowledge of the position of $\omega_{\rm knee}$ we 
divide the noise periodogram, $\tN(\omega)$,
 into logarithmically equally spaced bins below $\omega_{\rm knee}$ 
and linearly equally spaced bins above $\omega_{\rm knee}$.

Equations~\ref{eq:maxLmap} and \ref{eq:maxLnoise} are nonlinear and it is
therefore impractical to obtain an explicit solution for $\bN_\alpha$
and $T_p$. However the structure of this system lends itself to an
iterative approximation scheme; given the $i^{th}$ estimate the map,
$T^{(i)}_p$, we can find the the $i^{th}$ estimate of the noise in a
band $\alpha$, ${\bN}_\alpha$ from Equation~\ref{eq:maxLnoise}. As we
shall see, this system will converge to the MLE of $T_p$ and
$\bN_\alpha$. We note that this algorithm is similar to
``Expectation-Maximization'' (EM) algorithms \cite{EMalgorithm} which
alternate between filling in unknown data (as in our mapmaking step) and
calculating a maximum-likelihood estimation of some quantity (the
power-spectrum estimation step).

The algorithm starts with initial conditions:
\begin{eqnarray}
  T^{(0)}&=&0 \nonumber \\
  \bN_\alpha^{(0)} &=& {1\over n_\alpha+2}\sum_{\omega\in\alpha}
  |{\tilde d}(\omega)|^2.
  \label{eq:ic}
\end{eqnarray}
The iterative step is
\begin{eqnarray}
  T^{(i)}&=&\left( A^T N^{(i-1)-1} A \right)^{-1} A^T N^{(i-1)-1}d
  \label{it1} \\ 
  {\tilde s}^{(i)}(\omega)&=&{\tilde A}({\omega})T^{(i-1)}\nonumber
  \\
  \bN_\alpha^{(i)}&=&{1\over{ n_\alpha+2}}\sum_{\omega\in\alpha}
  |\td(\omega)-\ts^{(i)}(\omega)|^2.
  \label{it2}
\end{eqnarray}

The first model CMB experiment we will analyse is an idealized
small sky mapping experiment for which we shall adopt the name {\it fence}
scan  \cite{tegmapsa}: for the first half of the experiment the beam
is swept back and forward horizontally, shifting downwards until the
entire patch has been covered; for the second half of the scan the 
beam is swept back and forth vertically, shifting horizontally to
the left, again until the patch is completely covered.
The end result is a set of perpendicularly crossed linked scans.
The second model scan is a small sky version of the original
Planck Surveyor scan strategy, where great circles intersect at the two poles
(we shall call it a {\it poles} scan): The beam sweeps horizontally back
and forth along a square patch of the sky, drifting downwards until 
 the bottom of the patch when  it reaches the end of the scan; at the
edge of each horizontal sweep the beam returns to the same pixels
(one on the left and one on the right). The crosslinking is then simply at
the two end pixels (which play the role of the poles in the full-sky,
great circle Planck scan).
We consider a map with $10\times10$ pixels for the fence scan and
$8\times10+2$ pixels for the poles scan. We generate 56000 observations
per scan. The noise is generated by a stationary, Gaussian number
generator with periodogram given by Eq. \ref{eq:noiseper}
and we choose $\omega_{\rm knee}T_s$=0.5 where $T_s$ is the duration of
one horizontal sweep; we have found this to be typical values in
the BOOMERanG and MAXIMA experiments. 

\begin{figure}
\centerline{\psfig{file=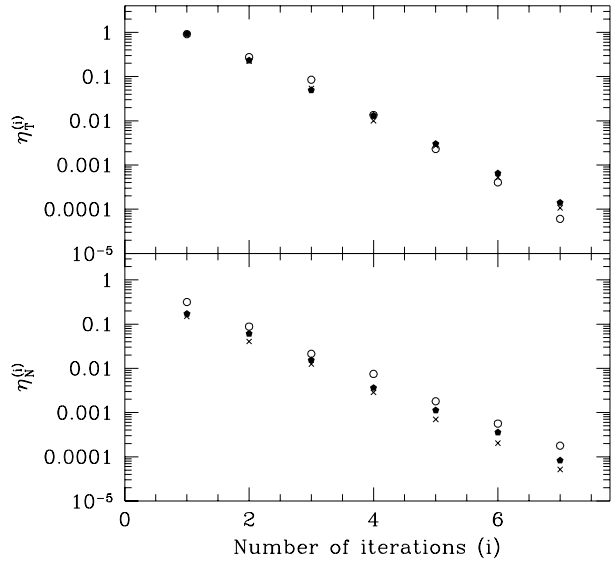,width=9cm}}
\caption{The convergence of the algorithm for the map, $\eta^i_T$ and
noise periodogram, ${\eta_N^i}$ for experiments with signal-to-noise
of 5 (open circles), 1 (closed pentagons) and 0.2 (stars).}
\label{fig:convergence}
\end{figure}

The first thing to check is if the algorithm converges. We quantify this
by comparing the map and periodogram in successive iterations using
\begin{eqnarray}
\eta^i_T&=&\frac{1}{N_{pix}}\sum_p
\frac{(T^{(i)}_p-T^{(i-1)}_p)^2}{(T^{(i)}_p)^2+(T^{(i-1)}_p)^2};\nonumber\\
\eta^i_N&=&\frac{1}{N_b}\sum_\omega
\frac{(\bN^{(i)}_\alpha-\bN^{(i-1)}_\alpha)^2}{(\bN^{(i)}_\alpha)^2
+(\bN^{(i-1)}_\alpha)^2},
\end{eqnarray}
where $N_b$ is the number of bins that characterize the periodogram
 and $N_{pix}$ is the number of pixels in the map. The form of
the denominator guarantees that spurious null values of these quantities
won't bias the measure of convergence. In Figure~\ref{fig:convergence} 
we plot the average of 200 simulated experiments for three levels of
signal-to-noise.

There are essentially only two parameters which may effect the existence
of, and convergence to, a fixed point: the number of samples per bin,
$n_\alpha$, and the signal-to-noise of the experiment.  As we saw in the
previous section, $n_\alpha$ must be sufficiently large for $C_{N{\rm
    eff}}$ to be positive definite; otherwise the problem is ill-posed.
As one would expect from Eq.~\ref{corrapp}, this is tied to the
convergence of the iterative scheme we propose: if the $n_\alpha$ are
small the algorithm either does not converge or takes a prohibitively
long time to converge (of order $10^2$ iterations). For simplicity of
analysis we consider all linear bins to have the same number of samples.
We then find that the threshold for both of the experimental strategies
is $n_\alpha\simeq2-5\times10^2$; the number of logarithmic bins between
the minimum frequency and $w_{\rm knee}$ are $N_<=({w_{\rm knee}/w_0})
({\ln n_\alpha}) / n_\alpha$.  For values greater than this, the system
has difficulties in converging, for smaller values there is
very little increase in speed.  A few comments are in order. Firstly in
choosing such values of $n_\alpha$ we are in a regime in which both the
joint estimate of $(T_p,\bN_\alpha)$ and the sole estimate of $T_p$
using the marginalized likelihood give essentially the same results.
Secondly one may worry that one is smoothing out important features in
the noise periodogram, in particular mis-estimating the shape of the
periodogram below $\omega_{\rm knee}$ and smoothing over spectral
``spikes''. We have found that in choosing $N_<$ logarithmic bins below
$\omega_{\rm knee}$ we adequately resolve the shape while at the same
time having enough samples per bin to get rapid convergence.  Fine
structures in the noise periodogram are typically wide enough to be
resolved with the resolution we propose here; very fine structures are
usually due to systematic effects and should be taken care of as such.
Thirdly, as we saw in Section~\ref{unknownnoise}, the wider the bins,
the smaller the correction to the noise covariance matrix; this must be
taken into account when subsequently using $C_N$ for estimating
$C_\ell$. For the ``borderline'' values of $n_\alpha$ we use for
Figure~\ref{fig:convergence} we expect the correction terms to increase
$C_N$ by $50$--$100\%$. Larger values of $n_\alpha$ for which
convergence is slightly more rapid should give smaller corrections.

The second factor we mentioned is the signal-to-noise of the
experiment; the figure of merit to be used is the signal-to-noise
per frequency band {\it in the bands where there is signal}.
Let us clarify what we mean. For the ensemble average
signal periodogram we know that there will only be a significant signal
amplitude for frequencies corresponding to the scan frequency
and its harmonics. Comparing to the ensemble average
noise periodogram, there will be two limiting regimes: if the
signal-to-noise is small, the noise will
dominate and the ``spikes'' of the signal will
be subdominant; if the signal-to-noise is large, the
signal periodogram, {\it at the scan frequency and its harmonics}
will stand out. The signal-to-noise in these bands is naturally
related to the signal-to-noise per beam; for the case of white noise
one has
\begin{eqnarray}
\left.\frac{S}{N}\right|_{\rm band}\simeq\left.\frac{S}{N}\right|_{\rm beam}
\end{eqnarray}
where the constant of proportionality is 
dependent on the characteristics of the
experiment (see Appendix~\ref{ston}).  The larger the S/N of the
experiment, the further away the initial estimate of the periodogram
will be from the ML estimate.  Prior expectation of the smoothness of
$\tN(\omega)$ can also be used to distinguish between noise and spikes
from the signal, although, as mentioned above scan-synchronous noise
(which produces similar spikes in the periodogram) is very difficult to
disentangle from the signal.  From Figure~\ref{fig:convergence} we can
read off the effect for different levels of signal-to-noise. For
$\eta_T^i$, we can see that, regardless of the level of signal-to-noise,
between $0.2-5$, the algorithm converges to better than one percent in
the first 5 iterations.  For signal-to-noise up to one, the speed of
convergence is approximately constant, while for high signal-to-noise,
after a transient of slow convergence, the map converges to the fixed
point much faster. The evolution of $\eta_N^i$ is much more sensitive to
the signal-to-noise: the higher the signal-to-noise the slower the
convergence at each iteration.

\begin{figure}
\centerline{\psfig{file=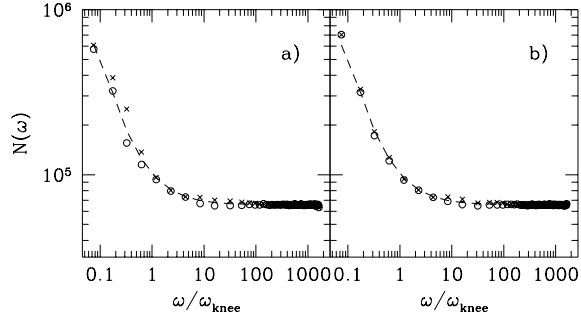,width=9cm}}
\vspace{-0.8in}
\caption{Results from 200 realization of the fence (a) and 
poles (b) strategy for signal-to-noise of 1. The dashed line is the
ensemble average periodogram, open circles are the average of the
MLE estimate after 5 iterations and crosses are the average of the
initial conditions of the periodogram as defined in Equation~\ref{eq:ic}
}
\label{fig:sn1a}
\end{figure}

\begin{figure}
\centerline{\psfig{file=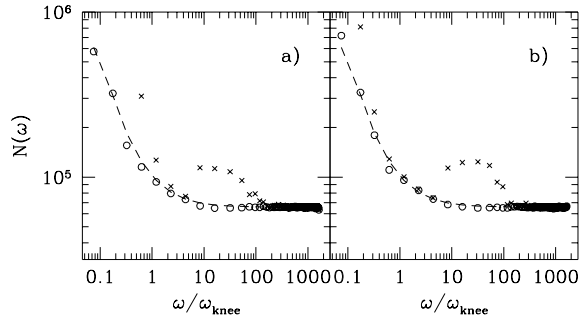,width=9cm}}
\vspace{-0.8in}
\caption{Results from 200 realization of the fence (a) and 
poles (b) strategy for signal-to-noise of 5. The dashed line is the
ensemble average periodogram, open circles are the average of the
MLE estimate after 5 iterations and crosses are the average of the
initial conditions of the periodogram as defined in Equation~\ref{eq:ic}
}
\label{fig:sn5a}
\end{figure}

Thus far we have shown that the algorithm converges to a fixed point
which we know is the MLE estimate of $T_p$ and $\bN_\alpha$. It is now
important to check how biased such an estimate is, \ie, whether for
an ensemble of realizations of the noise time series the average of the
MLE $\bN_\alpha$ is centred on the true ensemble average $\bN_\alpha$.
To do so we have generated 200 realizations of the mock experiments
for both fence and poles scans. In Figure~\ref{fig:sn1a} we plot the results
for an experiment with signal-to-noise per beam of order unity. The 
open circles are the MLE of the $\bN_\alpha$ after 5 iterations of 
our algorithm, and as we see match up with the ensemble average noise 
periodogram (dashed lines) for both the poles and fence scans. Moreover
we find that the initial estimate of the noise periodogram,
 $\bN^{(0)}_\alpha$ (crosses), is also a good estimate of the $\bN_\alpha$
at this level of signal-to-noise.

\begin{figure}
\centerline{\psfig{file=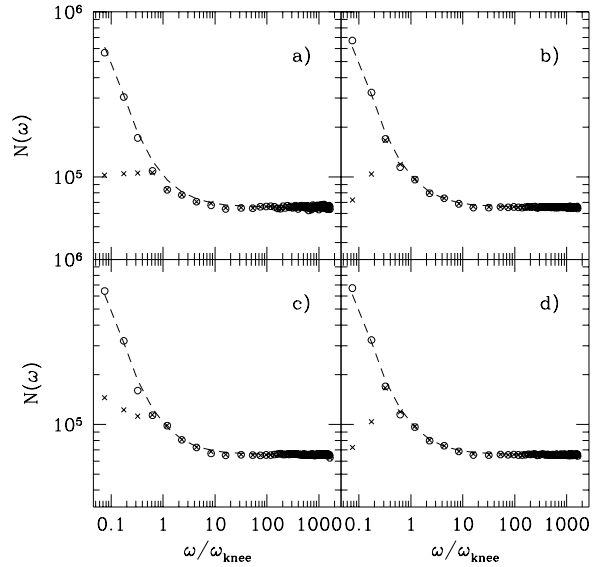,width=9cm}}
\caption{Results from 200 realizations of the fence (a) and 
poles (b) strategy for signal-to-noise of 1 and fence (c) and 
poles (d) strategy for signal-to-noise of 5. The dashed line is the
ensemble average periodogram, open circles are the average of the
MLE estimate after 5 iterations and crosses are the average of the
initial conditions of the periodogram as defined in Equation~\ref{eq:icraw}}
\label{fig:sn_raw}
\end{figure}

Naturally, for higher signal-to-noise per beam, the initial estimate
is not that good. In Figure~\ref{fig:sn5a} we plot the ensemble average $\tN(\omega)$
as a dashed line, $\bN^{(0)}_\alpha$ as crosses and the MLE of $\bN_\alpha$
as open circles for experiments with signal-to-noise of 5. In the high
signal-to-noise range of the periodogram, which corresponds to 
$\omega>\omega_{\rm knee}$, $\tN^{(0)}_\alpha$ is a very bad estimate of $\bN_\alpha$; one is misinterpreting signal as noise. We clearly see, however the
the MLE estimate of $\bN_\alpha$ is centred on the $\tN(\omega)$, a result 
which is not surprising given what we found in our analysis of the convergence
of the algorithm.

A possible alternative to initial conditions, $T^{(0)}_p$ and $\tN^{(0)}_\alpha$, of
Equation~\ref{eq:ic} is to start with the raw map:
\begin{eqnarray}
T_{\rm raw}^{(0)}=(A^TA)^{-1}A^Td.
\label{eq:icraw}
\end{eqnarray}
Indeed in the absence of low frequency, $1/\omega$ noise, this is the
MLE solution to the map. We have regenerated our monte carlo results
using these initial conditions and present them in
Figure~\ref{fig:sn_raw}. The first thing to note is that the MLE of
$\bN_\alpha$ is centred on the ensemble average $\tN(\omega)$ just as in
the original choice of $\tN^{(0)}$. Again this is evidence that the MLE
fixed point is a strong attractor of the iterative algorithm, and
relatively insensitive to initial conditions. In addition we find that
$\tN^{(0)}_\alpha$ constructed from the raw map estimate of the signal
is a bad estimator for $\bN_\alpha$ in the low frequency range. This is
true in both low and high signal-to-noise experiments given that at low
frequencies the noise component of the periodogram always dominates: the
raw map misrepresents the low frequency noise as signal and
subtracts it away from the data stream as such.

\section{Dipole Calibration}
\label{sec:dipole}
In this section, we apply the formalism developed in the earlier
sections to a somewhat different problem: the calibration of anisotropy
experiments off  the CMB dipole. Nowadays, many anisotropy experiments
have a ``spinning mode'' when they rotate in a circle of constant
elevation on the sky \cite{lee}. (Over short periods of time, this scan is
approximately a scan at constant ``latitude'' relative to some fixed
pole on the sky.) The circle is chosen to have sufficiently large
opening angle such that the measurement is sensitive to large-angle CMB
anisotropy.  The amplitude of the CMB dipole is sufficiently high that
the data for these scans is dominated by the modulation of the dipole
around the circle. In this case, the data are given by
\begin{equation}
  \label{eq:dipoledata}
  d_i = n_i + A_{ip}(\alpha_1 x_p + T_p + \alpha_2 g_p)
\end{equation}
where $n_i$ and $T_p$ are noise and anisotropy signal as before, and $x_p$
is the dipole pattern at pixel $p$, which has a well-measured amplitude
and well-understood pattern. The constant of proportionality for the
dipole, $\alpha_1$, is then the calibration of the detector.  We also
allow contamination from the Galaxy at that pixel, $\alpha_2 g_p$, where
$\alpha_2$ allows an extrapolation from the frequencies where the Galaxy is
well-measured to those observed by the experiment under consideration.
(Of course, if we take the signal, $T_p$, or the Galaxy, $g_p$, to be in
real temperature units, they too must include the calibration
parameter.)

We wish to basically apply the same rationale as before: determine
$\alpha$ along with the other parameters, or ideally determine $\alpha$
marginalizing over the other parameters.  For the purposes of dipole
calibration, we can ignore the intrinsic CMB contribution, or, for a
particular power spectrum, include it in the noise (because of the
circular nature of the scan, the isotropy of the CMB translates to an
effective stationarity for circular scans, so the CMB signal
contribution could easily be included). Because the signal-to-noise of
the dipole measurements is so high, it hardly matters. We can thus
rewrite the above equation in a suggestive form:
\begin{equation}
  \label{eq:dipole2}
    d_i = n_i + B_{i0} \alpha_0 + B_{i1} \alpha_1 = \nu_i + B_{ij} \alpha_j
\label{dip1}
\end{equation}
where $B_{i0} = A_{ip} x_p$ and $B_{i1} = A_{ip} g_p$. In this form, we
immediately see that we can use the techniques developed above to
estimate $\left\{\alpha_0,\alpha_1\right\}$. For known noise, the best-fit 
calibration would be
\begin{equation}
  \alpha_i = \left(B^T N^{-1} B\right)_{ii'}^{-1}
  \left( B^T N^{-1} d\right)_{i'} 
\label{dip2}
\end{equation}

In fact, if our model of the Galaxy were perfect, we could just set
$\alpha=\alpha_1=\alpha_2$ and estimate that single parameter: 
\begin{equation}
  \alpha = \left[(B_{i0} + B_{i1})^T N^{-1} (B_{i0} + B_{i1})\right]^{-1} 
(B_{i0} + B_{i1})^T  N^{-1} d 
\end{equation}
where now there are no matrix manipulations (except for $N^{-1}$).
However, our extrapolation of galactic models and observations remains
sufficiently uncertain that it is probably best to wholly ignore the
data within the galactic plane, which amount to a small number of pixels 
in a dipole scan.

With unknown noise, we can again iterate to simultaneously determine the
noise power spectrum and the calibration.  However, there is one
significant difference in this case: because the dipole scans describe
nearly perfect circles on the sky, the dipole signal is nearly
sinusoidal (and the Galaxy signal, although complicated, is nearly
periodic with the same period). Thus, most of the signal power will be
located in a very narrow frequency bin (or harmonics thereof) and will
have much higher signal-to-noise than the CMB anisotropy. This has
several practical effects. Outside of this bin, the noise power can be
determined by simply taking the periodogram of the data. If we expect
the noise power spectrum to be smooth compared to the width of the
dipole spike, we may be able to estimate the noise outside of the spike
and smoothly extrapolate into the region shared with the dipole. In this
case, we expect that the determination of the noise and calibration to
be nearly independent, and so the marginalization exercise above should
prove superfluous (as should any need for iteration).

The simplicity of this result is somewhat compromised by the
presence of the galactic signal.  To a very good approximation, we can
assume that the contribution from the Galaxy is zero except for a small
number of pixels where it overwhelms the rest of the data, and again
that the dipole signal is much larger than the anisotropy signal. As our 
models of moderate-latitude galactic emission improve, this information
can be incorporated. But because of the known angular frequency of the
dipole signal, it is safe to remove any part of the data possibly
contaminated by galactic emission.

In this case, the data are
\begin{equation}
  \label{eq:dipoledata2}
  d_i = n_i +\alpha A_{ip} x_p
\end{equation}
where we also take 
\begin{equation}
  \label{eq:dipolegalnoise}
  \langle n_i^2 \rangle \to \infty, \qquad
  \mbox{where~}\theta_i\in\mbox{galaxy}. 
\end{equation}
The formally infinite noise in these pixels just amounts to ignoring
those rows and columns in matrix manipulation. 

How does this method compare with more traditional methods of dipole
calibration? Rather then encompass all of the noise into a power
spectrum, it is customary to split the noise into
\begin{equation}
  \label{eq:dipolenoise}
  n_i = w_i + (a + b i)
\end{equation}
where $a$ and $b$ are constant over short segments of the data, and
$w_i$ is a white noise component. That is, the noise is described as an
offset, a linear drift, and white noise, over sufficiently short
timespans. The amplitude of the noise, $\sigma^2=\langle w_i^2\rangle$
is estimated and a least-squares fit to $a$, $b$ and the calibration is
done on the short segments and combined at the end of the process. This
can be complicated somewhat by marginalizing over the $(a,b)$ in each
segment and finding a single, global calibration and errors.  To the
extent that this is a good model for the noise, this procedure should
find correct results. Note, though, that Eq.~\ref{eq:dipolenoise} is not
a completely general description of noise while our result  simply
assumes Gaussian and stationary noise. 

Of course, this procedure can be extended to deal with any sort of
``template,'' that is, signals of known shape but unknown
amplitude. Such amplitudes can be estimated or even marginalized over in 
the usual way for Gaussian likelihoods. Here, though, we present it as a 
superior way to account for low-frequency noise when calibrating
experiments.

\section{Discussion \& Conclusions}
\label{sec:conclusions}
In this paper we have addressed the task of correctly estimating noise
in current CMB experiments. Our 
main points and conclusions are:
\begin{enumerate}
\item Long term, correlated noise is ubiquitous in current CMB experiments.
The dependence of the noise characteristics on the specifics
of the actual observational conditions make it essential to devise
an algorithm which can estimate the noise from the data stream itself.
\item Under the assumption of stationarity and Gaussianity of the
noise we present the Maximum Likelihood estimates of both signal
and noise from a data stream. These estimates are solutions to
a coupled set of nonlinear equations which must be solved iteratively.
\item Given that we are now jointly estimating the signal and noise,
there will be correlations between the map and the pixel noise
covariance matrix. The standard expression for the pixel noise
covariance matrix is now modified by a correction term which is 
effectively proportional to the number of pixels and inversely proportional to
the width of the bands one is using to
characterize the noise periodogram.  
\item If we forgo estimating the noise periodogram and marginalize 
over it, we obtain an estimate for the map and its covariance which
is effectively equivalent to the joint estimate of map and noise in the limit
of wide bands.
\item We show that, for wide enough bands, the iterative method
converges quickly to the fixed point. For convergence better
than a few percent, this can be achieved in 4-6 iterations,
for $S/N\simeq0.2-5$.
\item An analysis of two standard observational strategies shows
that our estimate is unbiased. We show that the two conventional
alternative methods are biased for high signal-to-noise.
\item It is conventional wisdom that an ideal observational strategy
has signal-to-noise of unity \cite{tegmapsa}. We show that, in this regime, a good approximation
to the estimator of the noise periodogram is the banded periodogram
 of the data stream. Thus one is able to avoid the time consuming process
of performing multiple iterations.
\end{enumerate}

A few additional comments should be made:
\begin{enumerate}
\item For Gaussian theories one can by-pass the map estimation step
  altogether, constructing an iterative algorithm that estimates
  ${\tilde N}(\omega)$ and $C_\ell$ directly. This sensibly puts the
  spatial power spectrum of the signal and the temporal power spectrum
  of the noise on the same footing.  A priori, however, the
  computational expense of such an algorithm is prohibitive; however, it
  is possible to simplify the matrix manipulations exploiting the
  stationarity of the noise and the isotropy of the signal. 
\item We have not addressed the computationally feasibility of our
  method; it is well known that for data sets of current CMB experiments
  and the expected megapixel datasets there is a serious problem with
  actually performing all the matrix manipulations
  \cite{borrill,delabrouille,wgh}.  In the case of our method these
  problems are of course worse: one must perform these matrix
  manipulations multiple times. We have limited ourselves
  to implementing our procedure assumimg the limitations of
  the standard map making procedure (we do not think that there
  is an alternative more efficient but accurate methods which
  can be applied for large data sets) and are therefore at the
 limit of computational capability with regards to data sets from
 experiments such as BOOMERanG and MAXIMA.
  A possible approach is to make local (in
  time) estimates of the noise periodogram by restricting oneself to
  subsections of the time stream which are long enough to show
  the long term correlations of the noise but small enough to be
  computationally tractable. This approach is indeed justified in some
  of the current experiments which have ``AC coupled detectors'' \ie,
  whose signal is high-pass filtered by the hardware. 
\item We have also not addressed the problem of ``gaps'' in the data
  \ie, the fact that sections of the data may be contaminated with, for
  example, cosmic rays, high amplitude galactic emission, fast detector
  transients, etc. There exist a variety of techniques for estimating
  the Fourier transform of unevenly sampled data, from the more
  conventional ones like the Lomb periodogram \cite{numrec} to advanced
  multitaper techniques \cite{imola}. For many data sets, only a small
  fraction of the time series is in gaps;  a simple filling-in procedure followed
  by a standard FFT periodogram will be adequate.
\item In Section~\ref{sec:dipole} we have only considered spatial
  templates (such as the Galaxy or the dipole). However, many CMB
  experiments suffer from the presence of signals which have a well
  defined structure in time \cite{kogut96}. Examples of such signals are
  detector variations due to the mechanical motion of the experimental
  apparatus or microphonic contaminants with well defined frequency and
  phase. Once again this can be dealt with in the way described in
  Section~\ref{sec:dipole} albeit with obvious modifications to
  Eqs.~\ref{dip1} and \ref{dip2} to include temporal templates.
\item For ground-based and balloon-borne experiments an essential problem
  is atmospheric contamination \cite{debernardis}.  The template is now
  spatially and temporally varying and therefore quite difficult to
  characterize in detail. A possible way of dealing with this
  contaminant is to have a channel dedicated to measuring atmospheric
  emission at high frequency during the time of observation; one then
  uses the data stream of this channel as a template. Of course this is
  an open problem and must be dealt with carefully in current CMB
  experiments.
\end{enumerate}

The techniques we have described here will be applied to the
MAXIMA \cite{lee} and BOOMERanG \cite{debernardisboom} experiments.

\section*{ACKNOWLEDGEMENTS}
We would like to thank J. Borrill, A. Lange, C.B. Netterfeld, J. Ruhl,
G. Smoot and especially S. Hanany and A. Lee for useful conversations.
Resources of the COMBAT collaboration (NASA AISRP grant NAG-3941),
NASA LTSA grant NAG5-6552, and NSF KDI grant 9872979 were used.

\appendix
\section{Analytical model of great circle scan}
\label{ston}

One would like to relate the signal-to-noise in the map to the signal-to-noise
in the time series. To do so we will work out a simple analytical
model of a one dimensional experiment \cite{dgh}. We shall consider a great
circle scan \ie,  choosing ${\bf x}_p=(\theta_p,\phi_p)$ with
$\theta=\pi/2$ and $0<\phi<2\pi$. If we assume a scan speed $\pdot=2\pi/T_s$ 
we can define
the pointing matrix to be
\begin{eqnarray}
A_{ip}=\delta[(i\delta t\pdot){\rm mod} (2\pi)-\phi_p].
\end{eqnarray}
The periodogram of the data stream is
\begin{eqnarray}
\left|\td(\omega_k)\right|^2=\left|{\tilde A}_{kp}T_p\right|^2+\tN(\omega_k)
\end{eqnarray}
where
\begin{eqnarray}
{\tilde A}_{kp}=\frac{2\pi}{\pdot}
\frac{\sin(\frac{n_s\pi\omega_k}{\pdot})}{\sin(\frac{\pi\omega_k}{\pdot})}
e^{{i(n_s-1)\pi\omega_k}/{\pdot}}e^{{i\phi_p\omega_k}/{\pdot}}
\label{eq:pointft}
\end{eqnarray}
and we have assumed that the total time of the scan is $T=n_sT_s$,
corresponding to $n_s$ consecutive scans of the great circle.

\begin{figure}
\centerline{\psfig{file=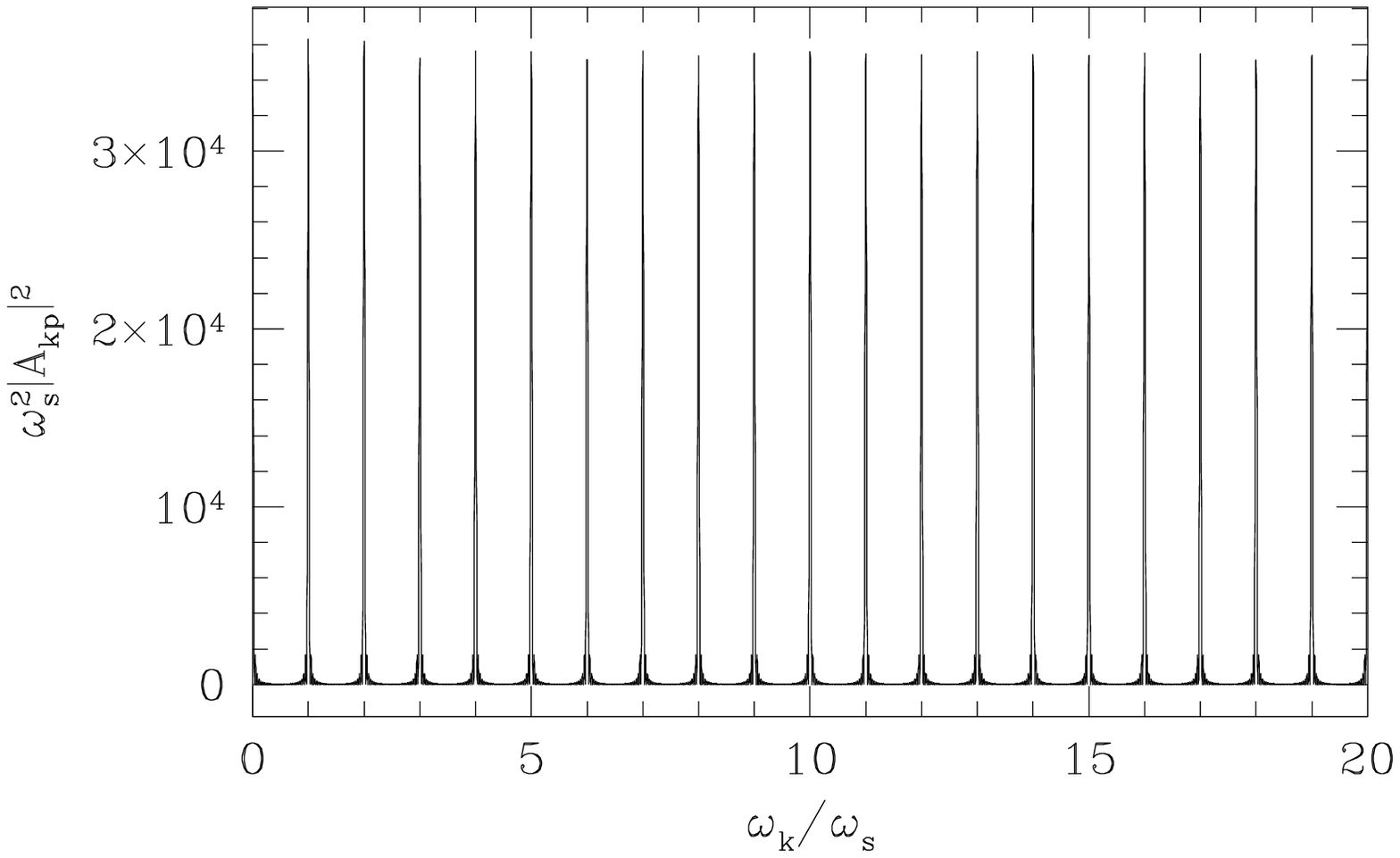,width=9cm}}
\caption{Dependence of $\omega_s^2|{\tilde A}_{kp}|^2$ on
$\omega$}
\label{fig:pointing}
\end{figure}

As we can see from Figure~\ref{fig:pointing},
$|{\tilde A}_{\omega\phi}|^2$ is highly peaked at multiples of
the scan frequency $w_s=2\pi/T_s$  (with peak amplitude $(n_sT_s)^2$);
this means that the signal in the time series is concentrated
at these frequencies. A natural measure of the signal-to-noise
in the time series is to compare the integrated power in signal for a band
around a multiple $k\omega_s$ of the scan frequency with the 
integrated power in noise
within the same frequency band, \ie,
\begin{eqnarray}
\left.\frac{S}{N}\right|_{\rm band}=\frac{\int_{(k-1/2)\omega_s}^{(k+1/2)\omega_s}d\omega
\langle|\ts|^2(\omega)\rangle}{\int_{(k-1/2)\omega_s}^{(k+1/2)\omega_s}d\omega
\langle\tN(\omega)\rangle}.
\end{eqnarray}
Naturally for arbitrary noise and signal, this ratio will depend
on the multiple of the scan frequency one is looking at. 
For example if the noise
has a $1/\omega$ the $\frac{S}{N}|_{\rm band}$ will be much smaller at low
$\omega$ than at high $\omega$. Let us consider a simple case where the
comparison is independent of $k$: 
$\langle T_pT_{p'}\rangle=\sigma_T^2\delta_{pp'}$ and 
$N_{tt'}=\sigma_N^2\delta_{tt'}$.
We then find that (looking at $k=0$):
\begin{eqnarray}
\int_{(k-1/2)\omega_s}^{(k+1/2)\omega_s}d\omega
\langle|\ts|^2(\omega)\rangle &=&2\pi n_sT_s\sigma_T^2 \nonumber \\
 \int_{(k-1/2)\omega_s}^{(k+1/2)\omega_s}d\omega
\langle\tN(\omega)\rangle &=& 2\pi\sigma_N^2.
\end{eqnarray}
The signal-to-noise in a given band is then
\begin{eqnarray}
\left.\frac{S}{N}\right|_{\rm band}= \frac{n_sT_s\sigma_T^2}{\sigma_N^2}=
\left.\frac{S}{N}\right|_{\rm beam}.
\end{eqnarray}

\end{document}